# Visual Observation and Quantitative Measurement of the Microwave Absorbing Effect at X band

L. Zhao[a], X. Chen, C. K. Ong

*Centre for Superconducting and Magnetic Materials, Department of Physics, National University of Singapore, Singapore 117542, Singapore*

**Abstract**

We have setup a 2D spatial field mapping system to measure graphically the quasi-free-space electromagnetic wave in a parallel plate waveguide. Our apparatus illustrates a potential application in characterizing the microwave absorbing materials. From the measured mappings of the microwave field, the visualization of spatial physical process and quantitative characterization of reflectivity coefficients can be achieved. This simple apparatus has a remarkable advantage over with conventional testing methods which usually involve huge, expensive anechoic chambers and demand samples of large size.

---
[a] Email: lzhao1027@ustc.edu.cn

# I. Introduction

Microwave absorbers have been long studied for the significance to the military application as the reduction of the radar cross section (RCS). Today with the proliferation of microwave telecommunication technology, Microwave absorbers are widely used in commercial marketplace to improve the performance and protect the core microwave circuit.

In order to asses the performance of different microwave absorbers, huge and expensive anechoic chambers are necessary to perform the quasi-"free-space" microwave reflectivity measurements [1]. Though this method can give a quick, repeatable non-destructive testing for the microwave absorbent materials over a wide frequency range, the spatial phase information is usually neglected. It was also difficult to extract experimental details about the interaction between the incident electromagnetic field and scattered one by the material under test..

Recently, the research group led by D. R. Smith in Duke University developed a point-by-point phase-sensitive method to obtain the spatial field mapping around metamaterials samples loaded in a two dimensional (2D) parallel plate waveguide [2]. Between two metallic plates, only the lowest transverse electromagnetic (TEM) mode can be excited, which resembles the electromagnetic wave in free space. The complex scattering problem of electromagnetic waves in the free space is reduced to 2D, so long as only scattering structures between two conducting planes do not break the translational symmetry along the z-axis (normal to the parallel plates). The rigorous

TEM mode can be preserved in which the electric and magnetic fields do not vary along the z-axis between the plates. Using their field mapping apparatus, they performed the first successful demonstration of metamaterial electromagnetic cloak at microwave frequencies(X-band) [3]. A similar experimental apparatus was lately developed by another group for the experimental verification about the overcoming of diffraction limit by a full-printed Veselago-Pendry lens [4].

Inspired by their pioneering work, we set up a similar field mapping system, and investigate its potential application in characterizing the microwave absorbing materials in this paper. Our system can not only visualize the significant effect of microwave absorbing coating on the propagating of microwave, but also give a very accurate quantitative test on its performance in a quite different way from traditional methods. It will be very meaningful to asses the absorbing performance using our simple measuring systems.

The details about the experimental setup and its performance test of are described in the following section. In Section III we demonstrated its important application in quantitative characterizing the microwave absorbers at X-band with different geometries. The results and corresponding conclusion are presented in the last section.

## II. Experimental Setup and Performance Test on Empty Chamber

The whole setup of our field mapping system is shown schematically in the Fig. 1(a). The basic geometry of the system is a parallel plate waveguide composed of two big aluminum plates, between which there are an 11mm air-filled spacing. The fixed larger top aluminum plate is 1.0m x 0.7m x 5mm, and supported by four pillars whose height can be adjustable respectively. The bottom plate is 0.6m x 0.4m x 3 mm, and mounted on a movable stage.

A coaxial detection antenna is inserted into (but not protruding below the lower surface of the top plate) the chamber through a small hole (with diameter about 2mm) in the center of the top plate. Its central probe-pin just flushes with the plate to reduce possible disturbance to the intrinsic field distribution in the chamber. An X-band waveguide adapter (Agilent Technologies, X281A , Type-N, 8.2 to 12.4 GHz) is fixed to the edge of the bottom plate as a source feed. The microwave electromagnetic fields introduced into the chamber are restricted between the two metallic plates. For air-filled parallel plate waveguides, the basic TEM mode has no cut-off while the other high-order modes, $TE_{mn}$ and $TM_{mn}$, have the same lowest cut-off frequency, $f_c=c/(2d)$ (c is the velocity of light in the air, and d the spacing of two conducting planes). So it is possible to excite only TEM mode inside the chamber, in which the electric and magnetic fields don't vary along the axis normal to the conducting planes (z-axis). The z-axis translation invariability will hold if the scattering objects inserted between the two plates don't break this symmetry. In our field mapping apparatus, the spacing between the two parallel plates is about 11mm. The cut-off frequency $f_c$ for the lowest order TM or TE modes is about 14GHz, consistent with the our current X-band feed source.

The bottom aluminum plate is mounted on a computer controlled XY-stage driven by two step-motors so that it can move freely in two dimensions relative to the fixed top plate. The XY stages can travel more than 200mm in two orthogonal directions respectively with the high resolution (<0.05mm).

A high sensitive vector network analyzer (VNA) are used (Agilent Technologies. type: HP8722D). Two flexible low-loss coaxial cables (Megaphase Inc. type: TM18) are used to connect the VNA ports to the adapter and antenna. The VNA provides the source microwave signal to the waveguide adapter and detects the return signal from the coaxial detection antenna, i.e. the complex transmitted parameter S21 is measured at each different position of the antenna. Because of the z-axis translational invariance, the electromagnetic field obeys the 2D scalar wave equation. The measured S21 reflects the corresponding local complex electric field directly,

$$E(x,y) = |E(x,y)|e^{j\theta} = \text{Re}[E(x,y)] + j\,\text{Im}[E(x,y)] \qquad (1)$$

The mapping range is enclosed by 10mm-thick saw-tooth-like microwave absorbing foams, and leaves an inlet for the incident microwave, which prevent any back reflection from the boundaries of the chamber, and suppress the possible resonant modes in the chamber.

The photograph of our final field mapping system is presented in the Fig. 1(b). The automated scanning and data acquisition process is controlled by a computer via both GPIB and RS-232 interface. The S21 parameters at all positions are stored in a complex matrix that can be plotted real-timely as 2D intensity graphs to visualize the electric field distribution at some specific frequencies.

To evaluate the performance of our measuring system, we first scaned the empty chamber without any scattering objects between two parallel plates(except absorbing foams at the boundaries). The scanned region is about 210mm x 210mm, about 180mm away from the outgoing aperture of the source feed at the edge of the bottom plate. The scanning step is 1mm.

The typical measured distribution of the real part of the electric field (corresponding to the real part of S21, marked by Re[S21]) at 10GHz is as depicted as 2D intensity graph in the left of Fig. 2. Microwave beams projected into the sample region from the lower side of the mapping region.

The graph vividly demonstrates the propagation of the electromagnetic waves in the positive y-direction. The space between the adjacent the phase fronts is 30mm, which is exactly the corresponding wavelength at 10GHz, as expected.

In our experiments, the field mapping can be acquired with highly repeatability, which indicates the good performance of our measure system. Because of the small aperture of the feed waveguide adapter and finite distance to the mapping region, the incident microwave is actually quasi-cylindrical-waves-like, which is manifested by the small curvature of the wave fronts shown in (a). The central area region can still be a good approximation to the case as illuminated by the plane wave in free space. To achieve a better simulation of a plane wave and extend the working region, it is necessary to adopt a pair of bigger parallel plates to let the mapping region far away from the aperture.

In order to quantitatively analysis the character of the empty chamber, we scan at 10GHz along the AB line, which is marked in Fig.2 (a). The measured real part (red triangle) and magnitude (black square) of the field are both shown in Fig. 2(b), indicating a quite good signal-to-noise ratio. The electric field exhibits the character of a decaying sinusoidal wave with the wavelength of 30mm. The slightly decaying of the magnitude comes mainly from the transverse expansion of cylindrical waves during the propagation, also partly from the metallic plates with finite conductivity and surface roughness.

Therefore, we can fit the measured data by an inverse-ratio decaying sinusoidal function $E = E_0 r^{-1} e^{j\beta r}$ ($\beta=\omega/c$ is the propagation constant). As shown in Fig. 2(b), there is a very good agreement between the measured and fitted data. These fitted parameters will be very useful when we carry out the subtraction of the decaying background for data analysis in further experiments.

## III. Experiments on Metallic Objects With and Without Absorbing Coatings

our field mapping system can provides a powerful tool to study scattering problem of the traditional materials. In this paper, our goal is to test the capability of our new field mapping apparatus in characterizing the microwave absorbers.

At present, almost all standard methods used to assess microwave absorbers require large and expensive anechoic chambers to achieve the quasi-free-space condition. Furthermore, it is necessary for the samples under test to be large sized. In our field mapping experiments, the only restriction on samples under test is that their height must be about 10mm to fit the spacing between the two parallel aluminum plates. A very small separation (less than 1 mm) is left to avoid dragging the top plate during the scanning process, for there is some permissible error on the parallelization of two aluminum plates in our homemade apparatus.

To characterize absorbing materials, we adopt a two-step procedure routinely. In our present work, the absorbing samples under test are 1.5mm-thick sheets made of magnetically loaded silicone. It is initially designed to have optimal absorption around about 10GHz. First, metallic objects such as bare aluminum bars or cylinders are inserted as a reference target placed in the center of the scanning region. The corresponding field mapping measurement is performed. In second step, we coated the reference objects with the microwave-absorbing sheet. The coating is adhesive to the surface of the metallic surface facing the incident waves. Then we return the coated samples back at their original positions in the chamber. The field distribution again while keeping. The comparison of results from two scanning experiments will give the performance of the absorbing coatings.

In Fig. 3, a bare aluminum bar of 100mm x 10mm x 5mm (marked by the gray region) is placed in the centre region with the largest face confronting the incident microwaves. The mapped field distribution, including the real part (left) and the magnitude (right) at 10GHz are depicted in Fig. 3.

Compared with the results for the empty chamber (shown in Fig 2a.), the incident electromagnetic waves are strongly reflected. The interference of the reflected and incident waves leads to the diffraction patterns of the wave fronts shown in (a). The aluminum bar reflects back most of the incident energy, leaving a shadow region behind the bar, which is more explicitly demonstrated in Fig. 3(b) (the magnitude plot). The period of strongly modulated magnitude in front of the bar is exactly half of the wavelength, which is a typical characteristic of standing waves.

We carried out to further quantitative analysis of our result by re-scanning along the AB line (shown in Fig 3). To acquire a higher signal-to-noise level, the center pin of the detecting coaxial antenna extended about 1mm into the chamber. The signal level increases nearly one order in comparison with measurements with the center pin not extended in the chamber. The drawback is the limitation of the range to avoid possible damage of the antenna pin by the sample during the scanning process.

The decaying background was subtracted from the raw data, using the fitting parameters for the empty chamber (Fig. 2(b)). Line scans of the real part, imaginary part, and the magnitude (c) of the measured fields along AB are shown in Fig.4. The decaying background has been subtracted and the normalized data are presented in Fig. 4. The reflected wave from the aluminum bar lead to standing wave and total electric field can be written as,

$$E(r) = E^+ e^{-j\beta r} + E^- e^{j\beta r} = E^+(e^{-j\beta r} + \Gamma e^{j\beta r}) = E^+ e^{-j\beta r}(1+|\Gamma|e^{j\theta}e^{j2\beta r}) \quad (2)$$

Where $\beta=\omega/c$ is the propagation constant for the parallel plate waveguides as well as the free space. $\Gamma = E^+/E^- = |\Gamma|e^{j\theta}$ is the reflection coefficient.

The standing wave ratio (SWR) is very useful to measure the reflectivity in many traditional microwave technologies. It is defined as

$$\text{SWR} = \frac{E_{max}}{E_{min}} = \frac{1+|\Gamma|}{1-|\Gamma|} \quad (3)$$

From our results shown in Fig. 4, the SWR along AB line is 14.8. Then the corresponding magnitude of $\Gamma$ can be calculated,

$$|\Gamma| = \frac{SWR-1}{SWR+1} \approx 0.87$$

In our measurement, both real and imaginary parts of field were acquired as shown in Fig 4. Therefore we can also obtain the phase of Γ by fitting our results according (2). We got $\theta = -146^o$ and $\Gamma = -0.829 - 0.559j$. Theoretically, the Γ should be -1 for perfect metallic reflector. However, there exists a very small air gap (about 1mm) between the top plate and the inserted samples during the scanning process. So a small quantity of incident electromagnetic energy is 'leaked' to back shadow region and |Γ| always less than unity. On the other hand, the air gap also made a capacitive contribution to the phase of Γ, result in the deviation from $180^o$.

In the second step, we coated the front face of the same aluminum bar with the 1.5mm-thick microwave absorbing sheet under test(marked as black section of line in Fig. 5). Fig. 5 depicts the field distribution of measured field in the chamber. The position of aluminum bar and the absorbing coating is also shown.

Comparing these field maps in Fig. 2, 3 and 5, we can see the striking effect of absorbing layer, which absorbed nearly all the incident energy without obvious reflection. The wave fronts of the microwave field before the cylinder remain nearly unchanged after inserting the sample into the empty chamber. In the geometrically shadow region behind the cylinder, the corresponding magnitude of the field reduces sharply. In the front region of the coated aluminum bar, there are still some weak

"ripples" in the homogenous background, indicating the existence of small standing wave due to minor reflection from our imperfect microwave absorber.

For further analysis, we carried out a line scan along the AB line shown in Fig. 6. The data processing procedures were the same as in previous step. The normalized real part, imaginary and magnitude of the electric fields are shown in Fig 6, where the decaying background has been subtracted. Compared with Fig. 4, the oscillations in magnitude are greatly suppressed, indicating the much smaller SWR due to the absorbing effect of the coating. Then we retrieved the complex reflection coefficient $\Gamma$ at 10 GHz as,

$$\Gamma = 0.17 e^{-j168^{O}} = -0.167 - 0.035 j.$$

So, our small simple apparatus can give not only the visual picture of the microwave scattering, but also quite good quantitative results. The error comes from the air gap between the sample and the top aluminum plates, also from the slight sagging of the two aluminum plates during the scanning process. The improvement in our future studies will involve the replacement of more rigid metallic plates and more accurately adjustment of the spacing to fit the samples with high repeatability and reproducibility.

Besides the above aluminum bars, many other metal objects of different shapes and incident configurations were also tested in our field mapping apparatus, without and with the same absorbing coatings. Some representative results of field distribution are shown in Fig. 7 and 8. Those manifestations visualize the remarkable effect of the microwave absorbing coating on the reduction of the reflection. Detailed data analysis together with numerical simulations will be discussed in elsewhere.

## IV. Summary

We have setup a simple field mapping measure system to describe graphically the 2D quasi-free-space electromagnetic wave in a parallel plate waveguide at the X-band frequencies. Although the current prototype system is sill very primitive and much improvement work will been done in the future study. Our apparatus illustrates a potential application in characterizing the microwave absorbing materials. The visual demonstration about the physical process and quantitative measurement of reflectivity coefficients can be achieved. This simple apparatus has a prominent advantage over with conventional testing methods which usually involve huge, expensive anechoic chambers and demand samples of large size.

## Acknowledgment

We would like to thank Dr P. Wang and Y. G. Ma for interesting discussions and offering absorbers in our experiments. And we are really indebted to the technical assistance from S. Sheng. This project also benefits from the support of the Defense Science and Technology Agency in Singapore.

**Figure Captions**

Fig. 1. (Color online) Schematic diagram (a) and photograph (b) of our 2D spatial field mapping system.

Fig. 2. (Color online) (a) Measured spatial map of the real part(left) of the electric field over a 210mm x 210mm region in a empty chamber at 10GHz. (b) Line scans of the measured real part(red triangle) and magnitude(black square) along the AB shown in (a). The dashed curves are corresponding fitted data.

Fig. 3. (Color online) Measured spatial map of the real part (a) and magnitude (b) of the electric field over a 210mm x 210mm region in a empty chamber at 10GHz. A bare aluminum bar (marked as gray region) located in the central region confronting the incident microwaves.

Fig. 4. (Color online) Line scans of the real part (red triangle), imaginary (green disk) and magnitude (black square) of the electric fields along the AB path shown in Fig. 3. The magnitude curve is re-plotted solely in the inset for clarify. The decaying background has been subtracted from all the raw data.

Fig. 5. (Color online) Measured spatial map of real part (a) and magnitude (b) of the electric fields at 10GHz over a range of 210mm x 210mm in the waveguide chamber. An aluminum bar (marked as gray region) is located in the center and its front face is coated with microwave absorber (black region).

Fig. 6. (Color online) Line scans of the real part (red triangle), imaginary (green disk) and magnitude (black square) of the electric fields along the AB path(shown in Fig. 5). The decaying background has been subtracted from the raw data.

Fig. 7. (Color online) Measured spatial map of the real part and the magnitude of the incident electric fields at 10GHz in the waveguide chamber in which scattered by an aluminum cylinder (with 20mm in diameter and 10mm in height marked as gray region) without((a) and (b)) and with ((c) and (d)) absorbing coating(marked as black ring). The scanning region is over a range of 210mm x 210mm

Fig. 8. (Color online) Measured spatial map of the real part and the magnitude of incident electric fields at 10GHz in the waveguide chamber in which scattered by an 10mm-high vertical aluminum bar (marked as gray region) without((a) and (b)) and with ((c) and (d)) absorbing coating(marked as black ring). The scanning region is over a range of 210mm x 210mm. The microwave is obliquely incident with the angle of about $45^{o}$ to the normal of the metal bar.

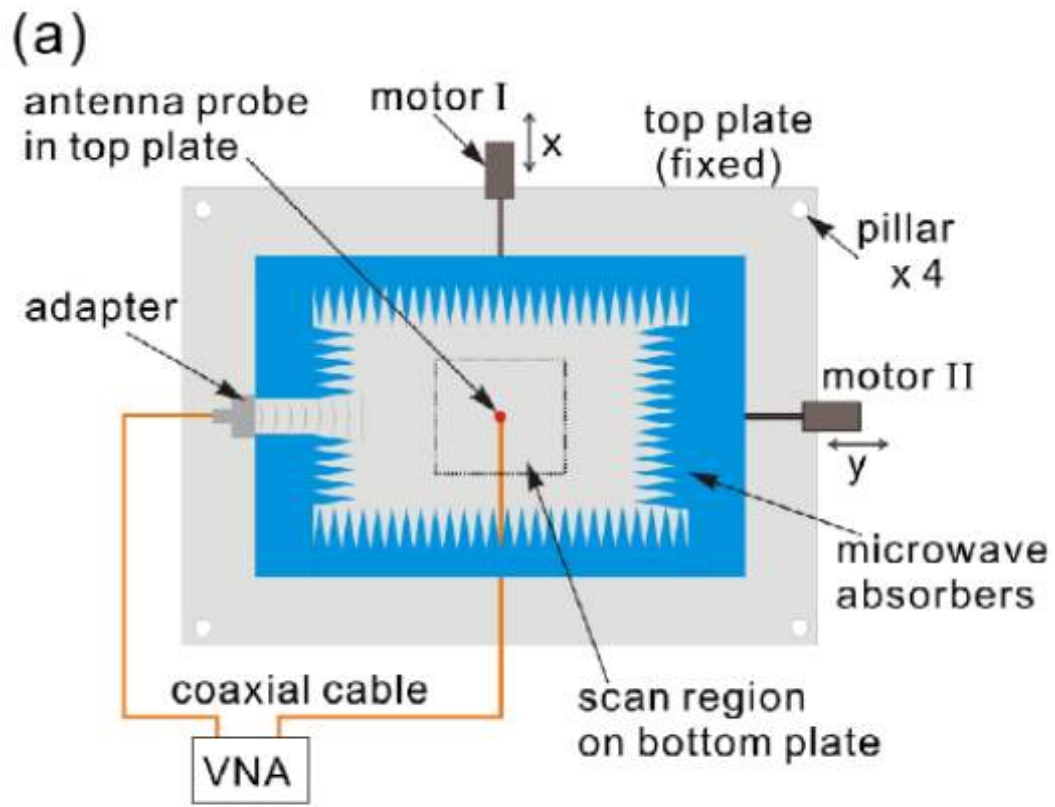
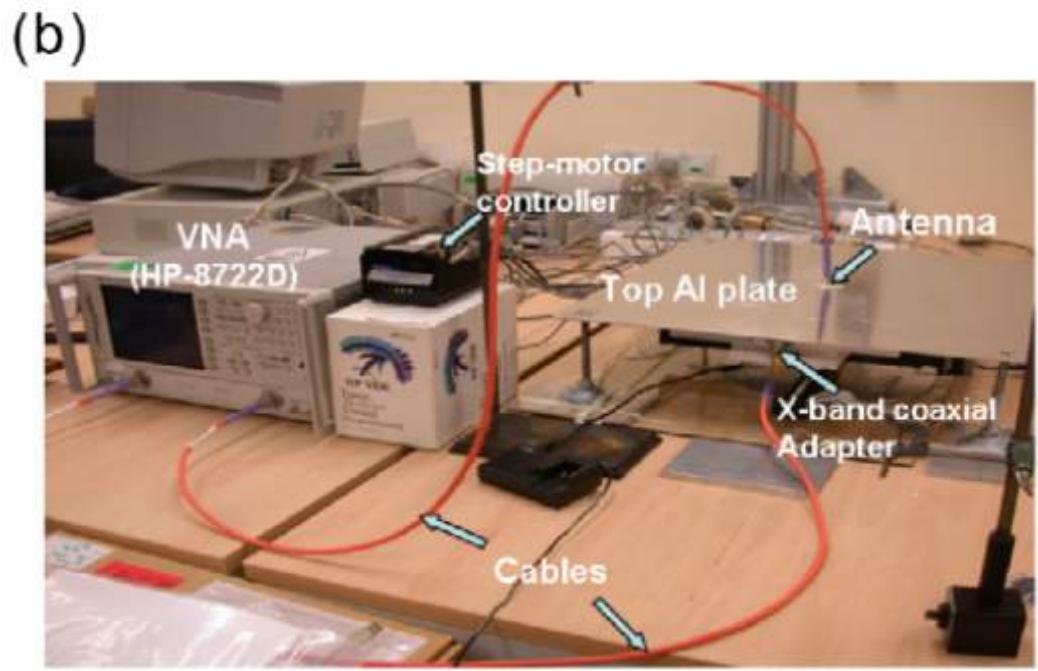

Figure 1

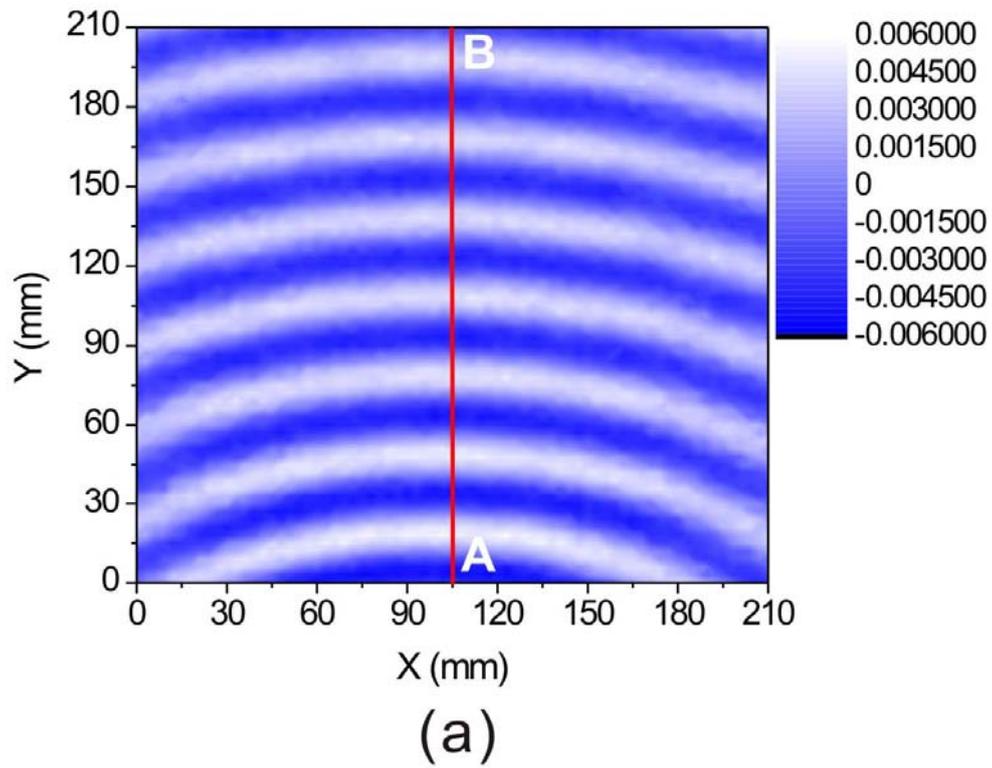

(a)

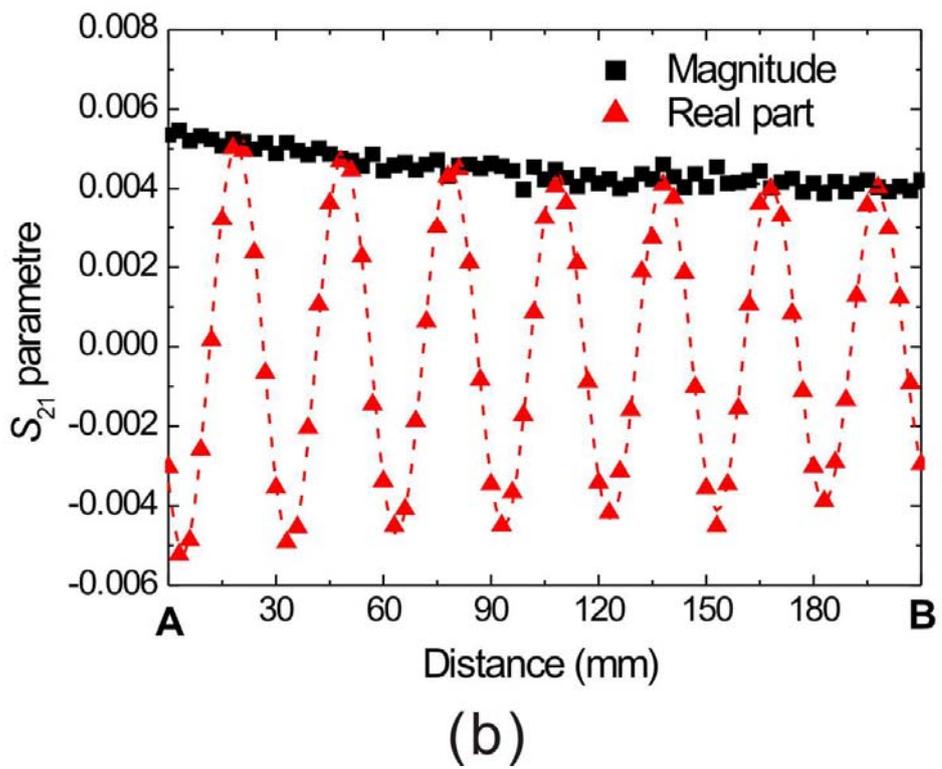

(b)

Figure 2

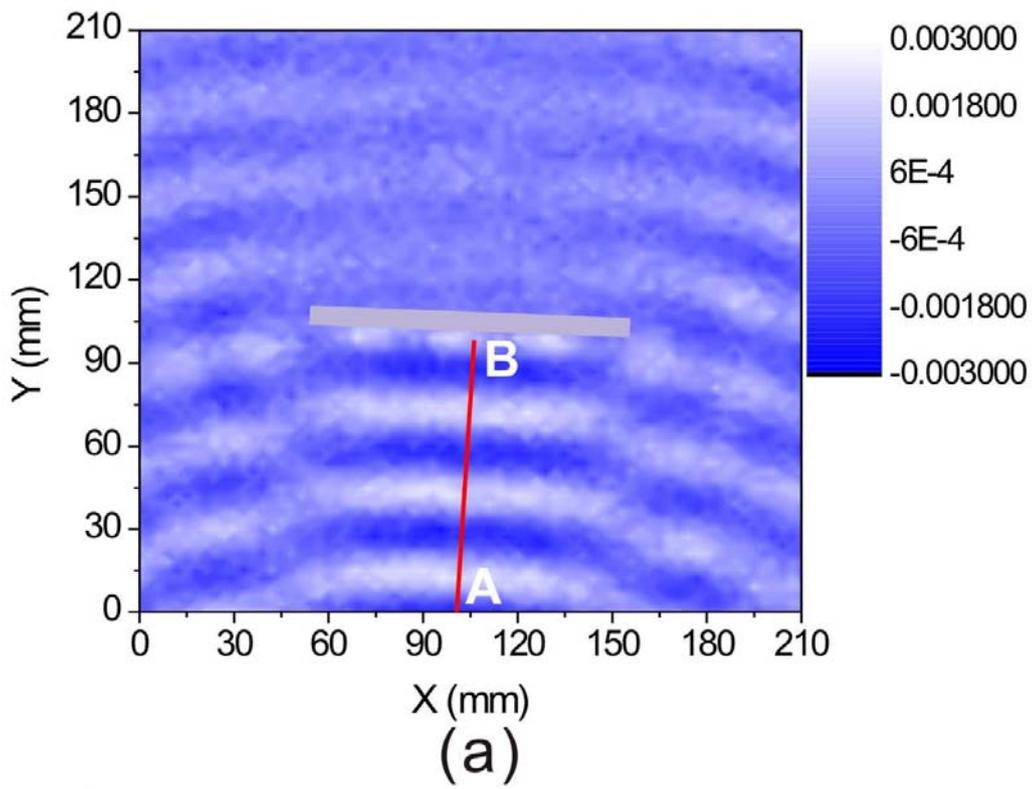
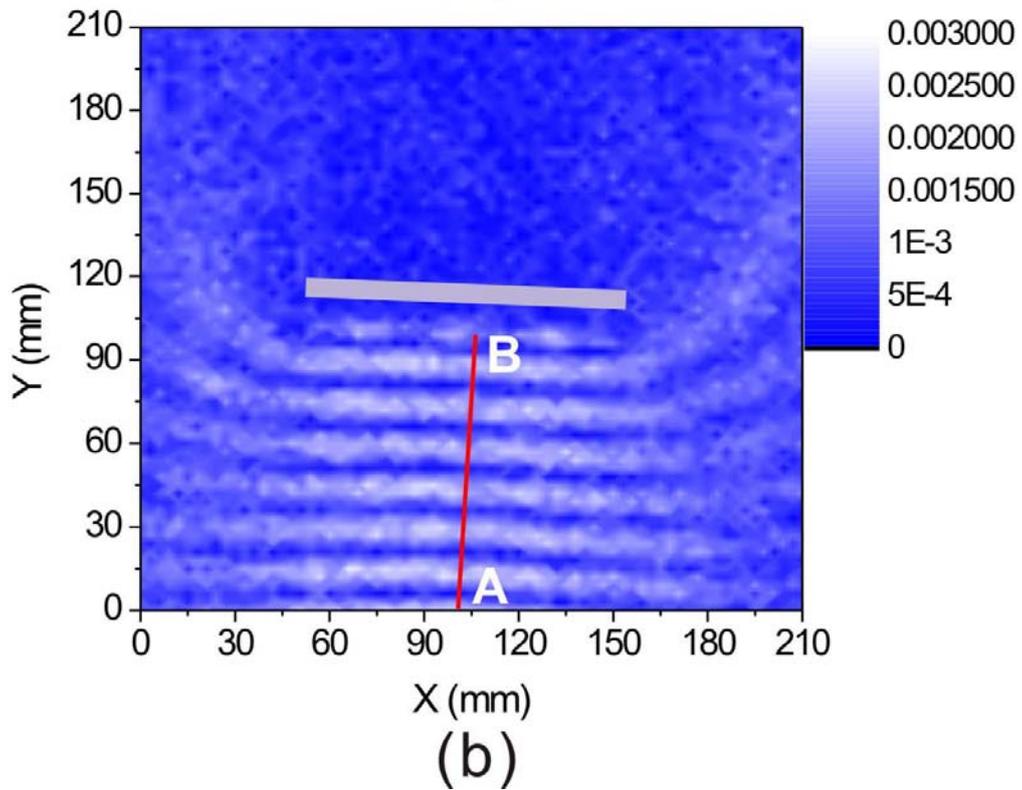

Figure 3

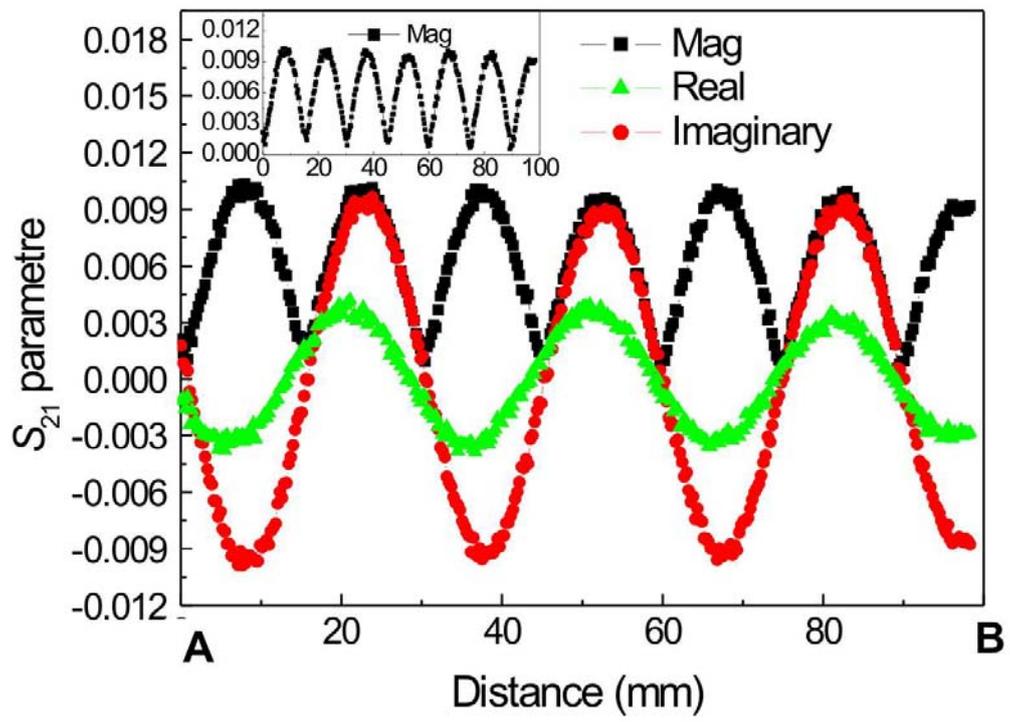

Figure 4

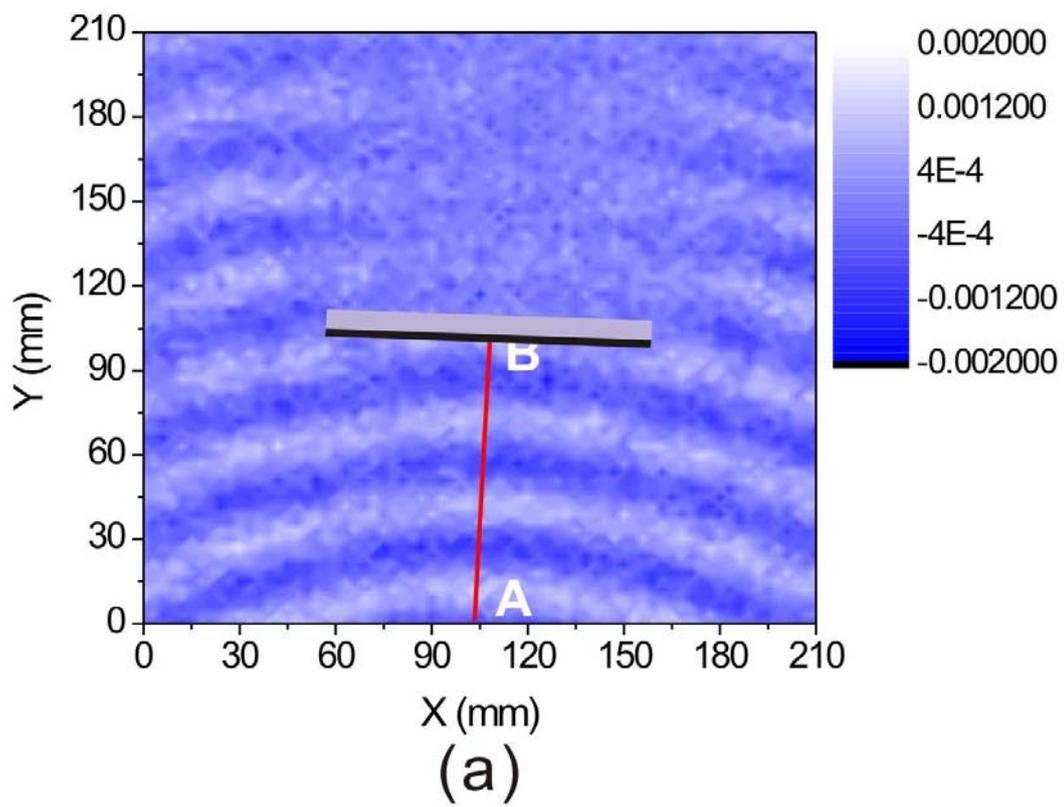

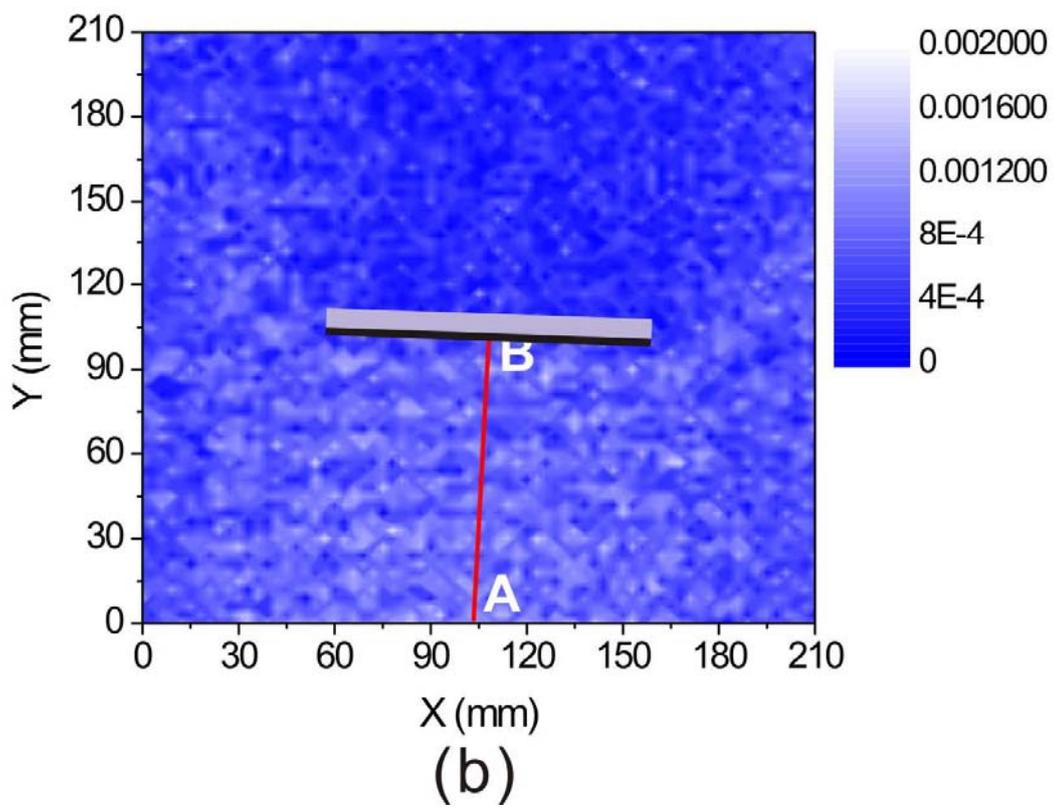

Figure 5

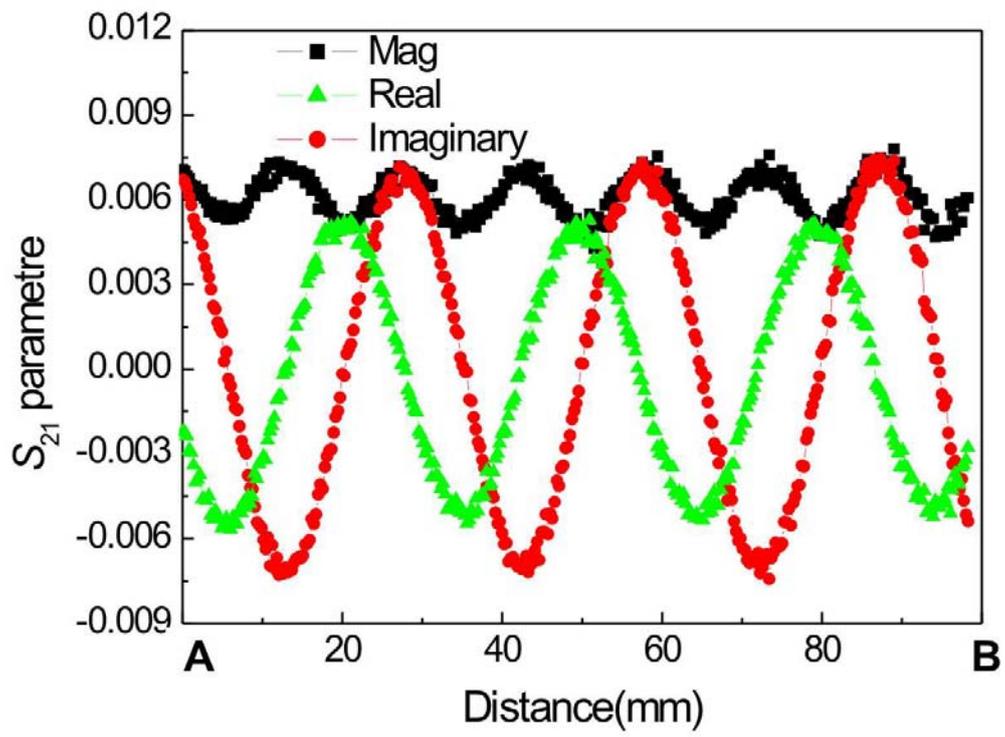

Figure 6

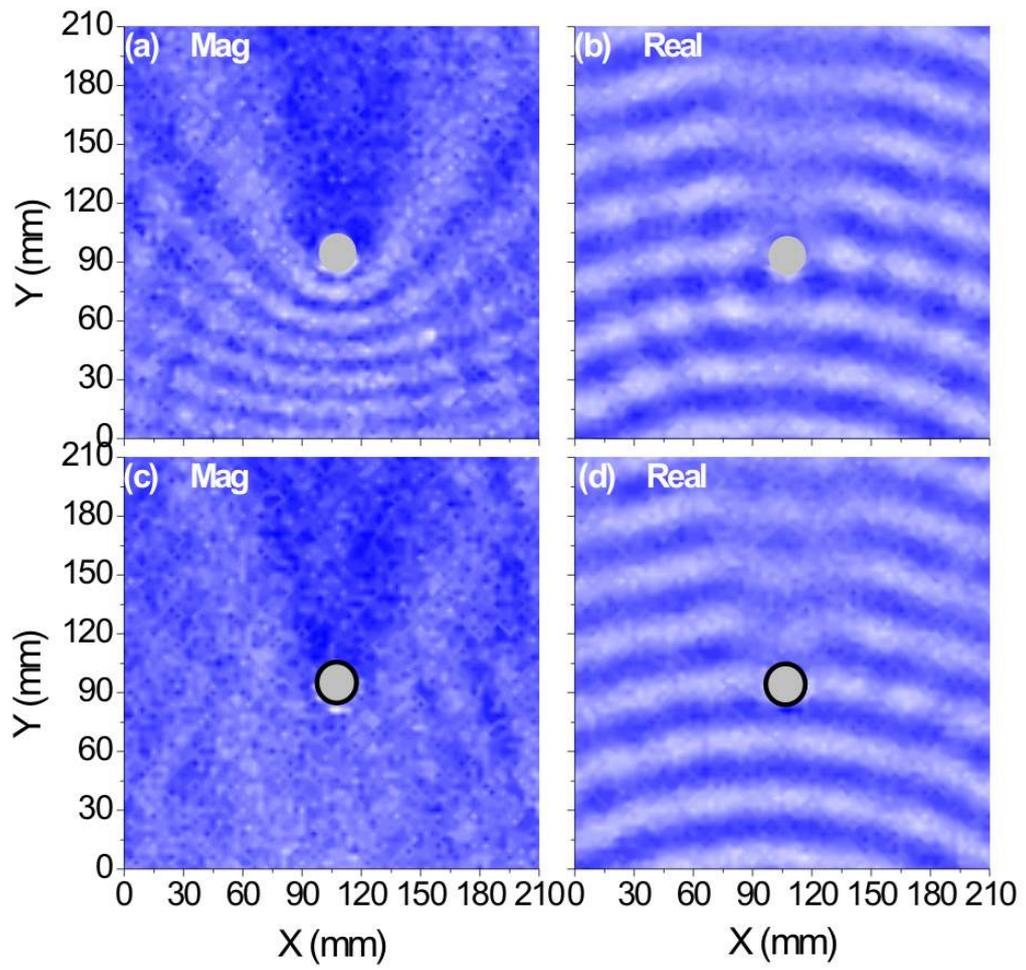

Figure 7

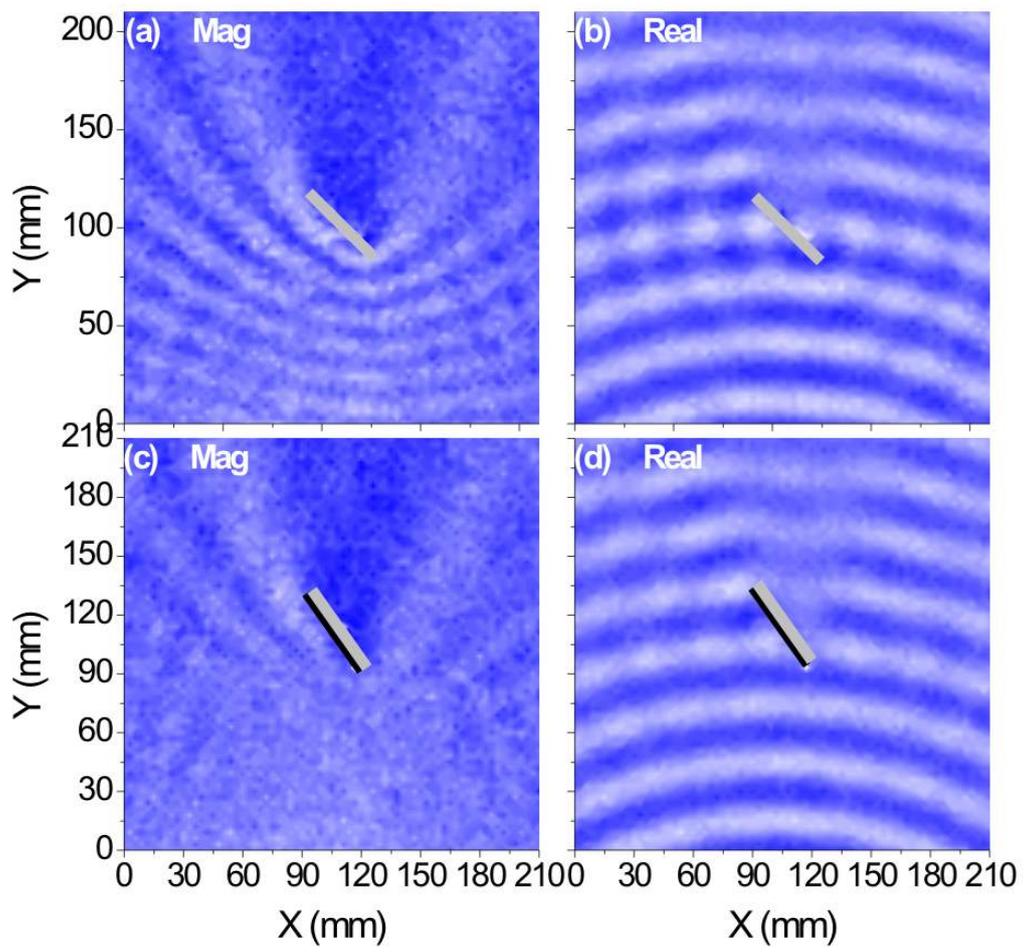

Figure 8